\documentclass[aps,prl,twocolumn,showpacs,citesort,tighten,amsmath,amssymb]{revtex4}
\usepackage{graphicx}
\usepackage{dcolumn}
\usepackage{bm}
\newcommand{\rd}{{\rm d}}

\begin{document}
\title{Asymmetric bubble disconnection:\\
persistent vibration evolves into smooth contact}
\author{Konstantin S. Turitsyn$^*$}%
\author{Lipeng Lai}
\author{Wendy W. Zhang}
\affiliation{%
Physics Department and the James Franck Institute
University of Chicago, Chicago IL 60637\\
$^*$Also at Landau Institute for Theoretical Physics, Moscow Russia}%

\date{\today}

\begin{abstract}
The disconnection of an underwater bubble illustrates how slight initial asymmetries can prevent the formation of a finite-time singularity.  Creating a singularity by focusing a finite amount of energy dynamically into a vanishingly small amount of material requires that the initial condition be perfectly symmetric.  In reality, imperfections are always present.  We show a slight azimuthal asymmetry in the initial shape of the bubble neck excites vibrations that persist over time.  As a result, the focusing singularity is generically pre-empted by a smooth contact. 
\end{abstract}
\maketitle

Whenever a diver plunges into a swimming pool or a waterfall hits a river, air is entrained, often as large cavities that subsequently break-up into many pieces~\cite{birkhoff57}.  Here we focus on a simple version of this disintegration process: the disconnection of an air bubble from a nozzle while it is submerged under water (Fig.~1a).  Previously it was believed that the bubble neck disconnects at a single point via an axisymmetric implosion that ends in a focusing singularity~\cite{longuet91,oguz93,gordillo05,burton05,keim06,bergmann06,thoroddsen07,egg07}.  In this scenario, water outside the bubble rushes inwards to constrict the neck of air, a motion converting the potential energy during the initial instant into kinetic energy.  Since the amount of water rushing inwards decreases to $0$ as the neck radius goes to $0$, all the kinetic energy is concentrated into a vanishingly small mass at the moment of break-up~\cite{schmidtpreprint}.  Analogous focusing singularities arise in models of sonoluminiscence~\cite{brenner03}, supernova and shock-wave implosion~\cite{zeldovich66}.  

Nearly all the previous studies analyzed the implosion by assuming that the focusing singularity controls the final dynamics.  However, Keim et al. found in experiments that a slight, azimuthal asymmetry in the shape of the bubble neck completely transforms the dynamics~\cite{keim06}.  An example of the different outcome is given in Fig.~1b.
Recently, Schmidt et al.~\cite{schmidtpreprint} showed that the initial asymmetry transforms the final disconnection by exciting vibrations in the cross-section shape of the bubble neck.  Once excited, the vibrations persist with the same amplitudes.  Since the average radius of the bubble neck is decreasing to $0$, the interface always becomes strongly distorted as break-up approaches. 
These constant-amplitude vibrations encode a detailed memory of the initial state. 
Such a detailed memory is surprising and important because the severely nonlinear evolution towards a singularity is commonly thought to be governed by convergence onto a universal dynamics, one independent of boundary and initial conditions and thus having little memory of the initial state~\cite{shi94,egg97,doshi03}. 
More generally, there are evidences supporting the view that 
such memory-encoding vibrations is a common feature of focusing singularities~\cite{schmidtpreprint, whitham57,plesset77,evans96,plewa04,maeda08,zwaan07,schmidtThesis}
\begin{figure}
\includegraphics[width=3.45in]{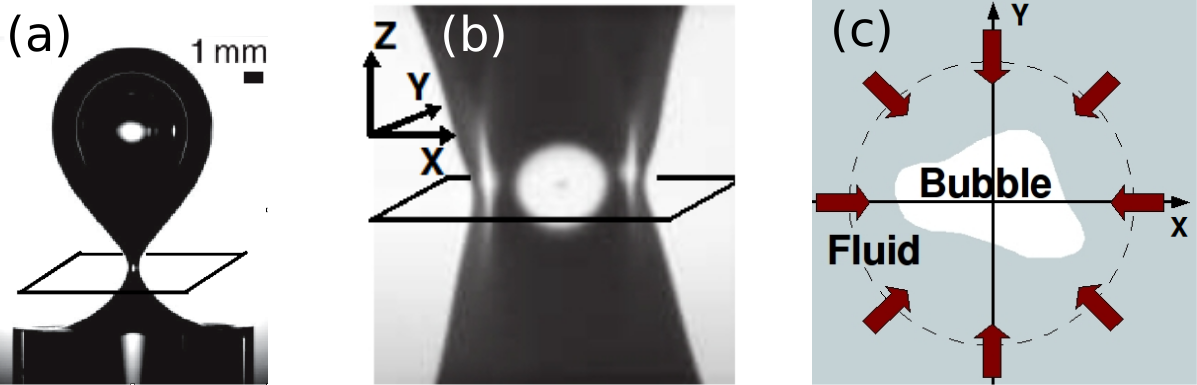}
\caption{Bubble disconnection dynamics.  (a) Experiment:  an air bubble (dark area) is submerged under water and released from a nozzle. Bright spots are optical artifacts. (b) When the bubble is released by a sudden burst of gas pressure from a slot, the initial ribbon-like neck disconnects via a "puncture", instead of breaking at a point.  There is a circular "hole" of water (light area) inside the ribbon-shaped bubble neck (dark area).  A satellite bubble (dark dot) is visible in the water-filled hole. See~\cite{keim06} for details. (c) Model: A cross-section of the neck (bubble region) contracts inwards due to a radially-symmetric influx of water from the far-field in the exterior.} 
\end{figure}

Here we focus on consequences of memory which cannot be addressed within the assumption of weak distortion employed by Schmidt et al~\cite{schmidtpreprint}.  We simulate the interface evolution in the final moment of break-up, when the surface is strongly distorted.  We find that the bubble neck generically evolves towards a finite-time contact.  Initially distant portions of the interface osculate, touching each other at a point and having a common tangent at the point of contact (Fig.~2b).  We also show that the finite-time contact can be viewed as a natural consequence of the vibrational amplitudes persisting with the same size while the average radius decreases to $0$. 

Our analysis starts from the simplest situation:  the break-up of a long and slender bubble neck.  Previous works have shown that, in this asymptotic limit, every cross-section of the bubble neck contracts solely due to the influx of water in the same horizontal plane, with no influence from the dynamics at other heights~\cite{longuet91,oguz93,egg07}.  We therefore focus on the cross-section at the height where the break-up will first occur.  The evolution of the cross-section at a fixed vertical height  corresponds to a two-dimensional (2D) version of the classic Rayleigh-Plesset collapse~\cite{landaubook} (Fig.~1c).  Stacking the 2D evolutions at the different heights vertically then generates the time-evolution of the entire neck.  We also neglect the effects of air flow within the bubble neck and surface tension at the interface.  Both effects can modify but do not change the essential features of our results.  Schmidt et al. have shown that the 2D model described above quantitatively reproduces the measured cylindrically-symmetric dynamics until the neck is a few microns across~\cite{schmidtpreprint}.   

We use a polar coordinate system whose origin coincides with the center of the neck cross-section. The air-water interface is given by $r=S(\theta, t)$.  Since viscous effects are negligible, the exterior velocity field is irrotational, i.e. $\mathbf{u}(\mathbf{x},t) = \nabla \Phi$ where $\Phi$ is a velocity potential.  Since the exterior flow is also incompressible $\nabla \cdot \mathbf{u} = 0$, the velocity potential $\Phi$ satisfies Laplace's equation $\nabla^2 \Phi = 0$.  At $t = 0$, a distorted bubble-neck, modelled as a region of constant and uniform pressure, is immersed in water.   To simplify the analysis, we choose the reference pressure level such that the bubble pressure $p$ is $0$. To drive the break-up of the bubble, we require that the exterior flow far from the bubble approach the form $(-Q/r)\mathbf{e}_r$ as $r \rightarrow \infty$.  
To simplify the calculations, we require that, $2 \pi Q$, the volume flux of the far-field in-flow, remains constant over time.  This is equivalent to requiring that $\bar{R}(t)$, the average radius of the bubble cross-section, decreases as $R_0 \sqrt{ (t_* - t)/ t_*}$, where $R_0$ is the initial value and $t_* = R^2_0/2Q$ is the time when the average size of the bubble goes to $0$.  This square-root decrease reproduces the leading-order behavior of the measured break-up dynamics~\cite{burton05,keim06,bergmann06,thoroddsen07,egg07,schmidtpreprint}.  Schmidt has shown that the higher-order, slow variation in $Q$ does not affect behavior of the vibrations at leading-order~\cite{schmidtThesis}. 
Two boundary conditions dictate the time-evolution of the velocity potential $\Phi(r, \theta, t)$ and the interface $S(\theta, t)$.  First, at the rapidly accelerating surface, the normal stress exerted by the exterior flow must equal the bubble pressure.  This unsteady version of the Bernoulli condition has the form 
\begin{equation}
[ \partial \Phi /\partial t + (1/2)|\nabla \Phi|^2 ] |_S = 0.
\label{eqn:bernoulli}
\end{equation} Second, the position ${\bf x}$ of a material point on the surface is advected by the exterior flow,\begin{equation}
\frac{\rd {\bf x}}{\rd t} = \left( \frac{\partial }{\partial t} + \nabla \Phi \cdot \nabla \right ){\bf x} = \nabla \Phi |_S
\label{eqn:kinematic}
\end{equation}

We use an approach developed by Dyachenko et al. to solve for the interface evolution~\cite{dyachenko96,zakharov02}.  First, the solution of $\nabla^2 \Phi = 0$ in the exterior is simplified by mapping the exterior of $S(\theta, t)$ conformally onto the exterior of a unit circle in the complex plane $w$ (see e.g. \cite{shraiman84}).  A point on the $w$-plane is related to the location $(r, \theta)$ on the 2D plane by $z(w,t)=re^{i\theta}$. The velocity potential in the physical plane is then given by the real part of a complex velocity potential $\Psi$. Second, instead of solving directly for the interface $z=S(\theta, t)e^{i\theta}$ and the complex potential $\Psi$, we work with the new variables 
\begin{equation}
{\cal R}(w,t) = 1/(w \partial_w z) \quad {\cal V}(w,t) = (\partial_w \Psi)/(\partial_w z) . 
\label{eqn:RV_def}
\end{equation}
The variable $\cal V$ corresponds to the speed of the fluid motion on the interface.  The variable $\cal R$ does not have a straightforward physical interpretation, though it clearly is a measure of how distorted the void shape in the real space has become relative to the unit circle on the $w$-plane. 
Equations (1) and (2) now assume the form
\begin{eqnarray}\label{Req}
&	\partial_t {\cal R} = w (\partial_w {\cal R}){\cal A}\left\{\mathrm{Re}[\cal R \cal V^*]\right\} - w {\cal R}\partial_w{\cal A}\left\{\mathrm{Re}[\cal R \cal V^*]\right\} \\
&	\partial_t {\cal V} = w (\partial_w{\cal V}){\cal A}\left\{\mathrm{Re}[\cal R \cal V^*]\right\} - w {\cal R}\partial_w{\cal A}\left\{|{\cal V}|^2\right\}/2\label{Veq}
\end{eqnarray}
where the $*$ symbol denotes complex conjugation. The integral operator ${\cal A}$ is the Cauchy integral (see e.g. \cite{carrier,shraiman84}):
In the simulation, we represent the initial state, i.e. the surface shape $S$ and the velocity potential $\Psi$, in terms of $\cal R$ and $\cal V$, as an expansion with $N$ total modes, 
${\cal R} = \sum^N_{n=0} {c_n}/{w^{n+1}}$ and ${\cal V} = \sum^N_{n=0} {d_n}/{w^{n+1}}$.
The coefficients $c_n$ and $d_n$ are computed from $\Psi$ and $S$.  We then update the values of $\cal {R}$ and $\cal V$ using (\ref{Req},\ref{Veq}).  The new neck shape $S$ and velocity potential $\Psi$ are in turn computed by numerically integrating (\ref{eqn:RV_def}).  
This formulation speeds up the numerical computation because the right hand side of equations (\ref{Req},\ref{Veq}) can be computed in $N \log N$ steps using the Fast Fourier Transform.  Results presented below have been obtained with $N=256$. Using $N=1024$ or more points yields no significant 
changes.

Fig.~2a and 2b give the outcome of a typical simulation.  The average radius $\bar{R}$ decreases as 
$\sqrt{(t_* - t)/t_*}$ (Fig.~2a).  As the bubble region shrinks, an initially slight distortion becomes more and more noticeable.  To illustrate how the distortion develops we have chosen $6$ moments (labelled in Fig.~2a) and rescaled them by $\bar{R}(t)$.  At $t=0$ the void shape is a nearly circular oval, given by $S(\theta, t=0)= R_0 + A_2 \cos \Omega_2 \cos(2\theta)$ where ${R}_0$ is the initial value of $\bar{R}$, $A_2$ the size of the azimuthal distortion and $\Omega_2$ the initial phase.  During the early moments, $A_2 \ll \bar{R}(t)$, so the cross-section vibrates, in quantitative agreement with linear stability results~\cite{schmidtpreprint},
\begin{eqnarray}
S(\theta, t) &=& \bar{R}(t) + A_2 \cos[\phi_2(\bar{R})] \cos (2\theta) \\
\phi_2 (\bar{R}) &=& \ln(R_0/\bar{R}(t)) + \Omega_2 .
\label{eqn:linear}
\end{eqnarray} 
The amplitude of the vibration $A_2$ remains constant while the phase of the vibration $\phi_2$ oscillates. In Fig.~2b, the cross-section is initially elongated along the $x$-axis, then along the $y$-axis, and later is elongated along the $x$-axis again.  As the break-up proceeds, the interface becomes strongly distorted ($A_2 \approx \bar{R}(t)$), evolving out of the the linear stability regime. The cross-section deforms into a dumb-bell shape. As time goes on, the waist of the dumb-bell narrows and its lateral extent broadens.  Finally, the opposite sides of the surface touch, with a common tangent at the point of contact, thus severing itself into two side-by-side lobes.  This is the typical outcome when an azimuthal asymmetry is present.  It is also qualitatively consistent with the "puncture" type break-up (Fig.~1b) observed in the experiment by Keim et al~\cite{keim06} if one assumes that the satellite bubble at the center of the "puncture" is created by viscous drag created by the drainage of air from the thin gap in the final instants of contact.  More complicated, but less frequent, types of self-intersection will be analyzed in a future study.  
\begin{figure}
	\includegraphics[width=3.45in]{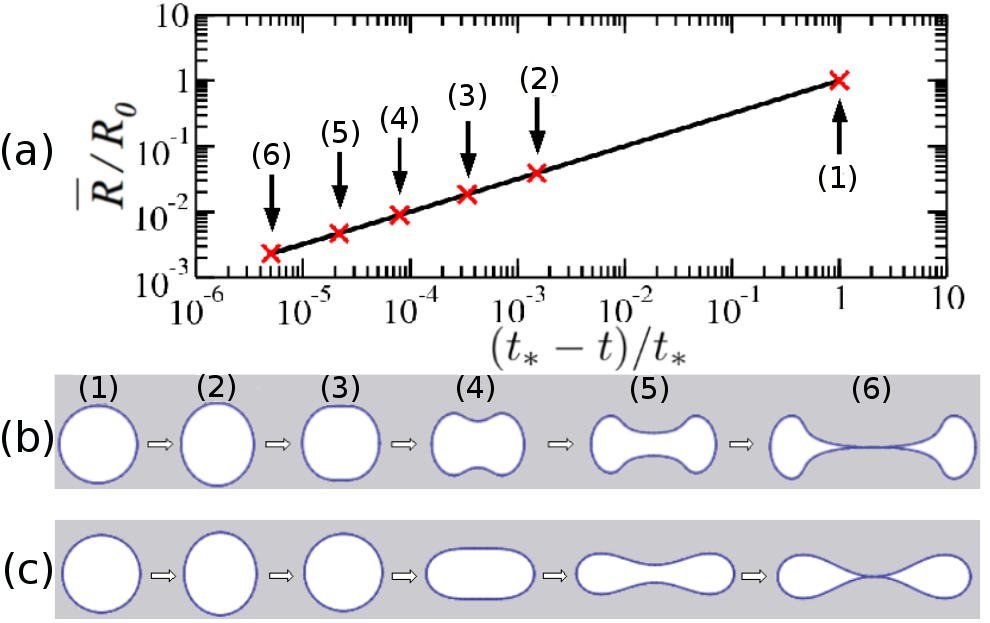}	
	\caption{Time-evolution of a slightly asymmetric neck cross-section.  The initial distortion is a single $n=2$ vibrational mode ($A_2 / R_0=0.004$, $\Omega_2 = \pi/4$). (a) The average radius $\bar{R}$, rescaled by the initial average radius $R_0$, as a function of $ (t_* - t )/ t_*$, the amount of time remaining until the point implosion singularity at $t_*$. (b) Neck cross-section at successive times.  Shapes rescaled by the average radius $\bar{R}(t)$. (c) Shape evolution when $n=2$ linear stability dynamics is extrapolated as a series expansion in $\cal{R}$, $\cal{V}$ variables down to contact (Eqn.~(8)).}
\end{figure}

Fig.~3 shows how $U_c / (\rd \bar{R}(t)/\rd t)$, the rescaled velocity at the contact location, varies as the dimensionless gap width $d/R_0$ goes to $0$.  Initially, while $d$ is large, the interface accelerates inwards faster than the average radius decreases, or $U_c /  (\rd \bar{R}(t)/\rd t)  > 1$.  As $d$ decreases, this rescaled velocity reaches a maximum and then slows.  Eventually the rescaled contact speed settles to a constant value as $d$ goes to $0$.  In Fig.~3, the simulation curves starting from $3$ different initial conditions clearly 
terminate with different contact speeds, showing that $U_c$ is chosen by the initial state.  Note also that 
the simulation typically yields $U_c / (\rd \bar{R}(t)/\rd t) \approx O(1)$, i.e. the cross-section shrinks about as fast as the two sides of the interface collides at the moment of contact. 
\begin{figure}
\includegraphics[width=3in]{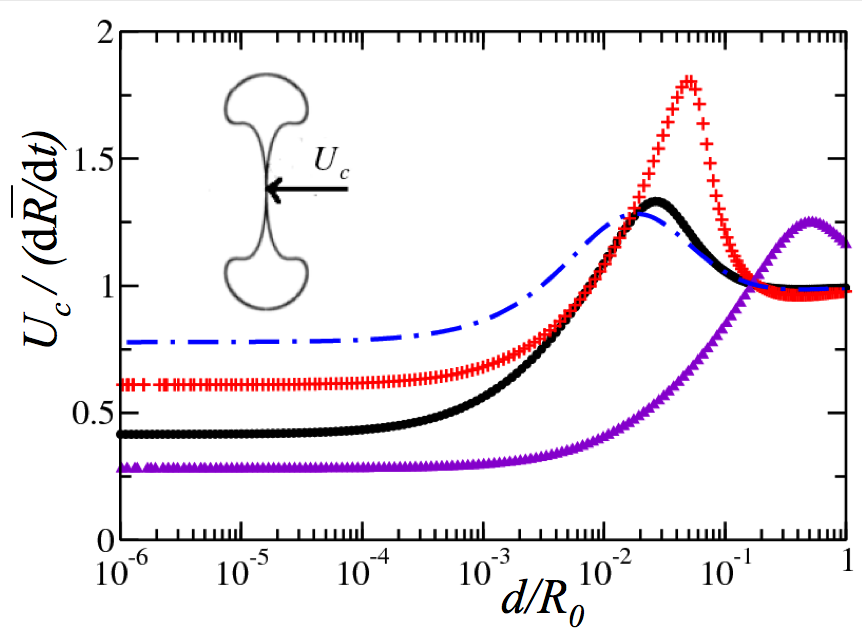}
\caption{Velocity at the contact location, $U_c/(\rd \bar{R}/\rd t)$, as a function of the gap width $d/R_0$.  Simulation results for $A_2/R_0=0.004$ (circles), $0.011$ (crosses) and $ 0.067$ (triangles) with the same $\Omega_2 = \pi/4$ display the same qualitative behavior but terminate with different $U_c$ values. The dot-dash line is prediction from Eqn.~(9) with $A_2/R_0 = 0.004$ (circles).}
\end{figure}

It is possible to give a simple interpretation of the qualitative features found in these simulations. 
The contact dynamics found in a typical run can be thought of as an "extrapolation" of the amplitude-preserving vibration dynamics excited in the linear stability regime. Specifically, we can rewrite the linear stability results for $n=2$ vibration (Eqn.~(6,7)) as a Taylor series expansion in $\cal R$, $\cal V$ (Eqn.~(8,9)).  
\begin{eqnarray}
{\cal R}(w, t) &=& \frac{1}{w \bar{R}(t)}\left[ 1 + \left( \frac{A_2\cos[\phi_2(\bar{R})]}{\bar{R}(t)} \right) \frac{1}{w^2} \right ] \\
{\cal V}(w, t) &=& \frac{-1}{w\bar{R}(t)}\left[ 1 + \left( \frac{A_2 \sin[\phi_2(\bar{R})]}{\bar{R}(t)} \right) \frac{1}{w^2 } \right ].
\label{eqn:calRseries}
\end{eqnarray}
As $A_2$ becomes comparable with $\bar{R}(t)$, Eqn.~(8,9) represents a particular scheme for 
approximating how the vibrations transform when nonlinearity is significant. For our problem, this scheme successfully reproduces all the qualitative features found from the full simulation.  This is evident  
in Fig.~2c, where the hypothetical shape evolution (Eqn.~(8)) predicts that the cross-section osculate and reproduces the successive shape changes. Similarly, the time-evolution of the speed at the contact location predicted by Eqn.~(9) show the same qualitative trend as the full simulation (circles), though the prediction does over-estimate the value of the final contact speed. Mathematically, this agreement says that $\cal R$, $\cal V$ are "natural" variables for tracking the break-up dynamics, since the time-evolution assumes such a simple form in $\cal R$, $\cal V$. Physically, 
this supports a simple scenario for how the initial distortion control the final contact dynamics:  when the initial distortion is dominated by a single vibrational mode, the final contact is dominated by the same mode.  Although the nonlinear interactions can and do generate new modes and cross-link different modes, in a typical situation neither mechanism gets enough time to cause qualitative changes.  
Fig.~4 plots $\bar{R}_c$,  the average radius of the cross-section at the moment of contact, as a function of the initial shape asymmetry $A_2/R_0$.  The prediction from Eqns.~(8,9) (lines) reproduce both the contact orientation and, roughly, the values of $\bar{R}_c$ found from the simulations. Overall, $\bar{R}_c$ decreases with $A_2$, since smaller initial asymmetry allows the axisymmetric implosion to proceed further. The step-like structure associated with successive changes in the contact orientation reflects the fact that the final dynamics is dominated by the initial $n=2$ distortion.  Since a contact due to $n=2$ vibration alone can only occur along two distinct orientations.  
\begin{figure}
\includegraphics[width=3in]{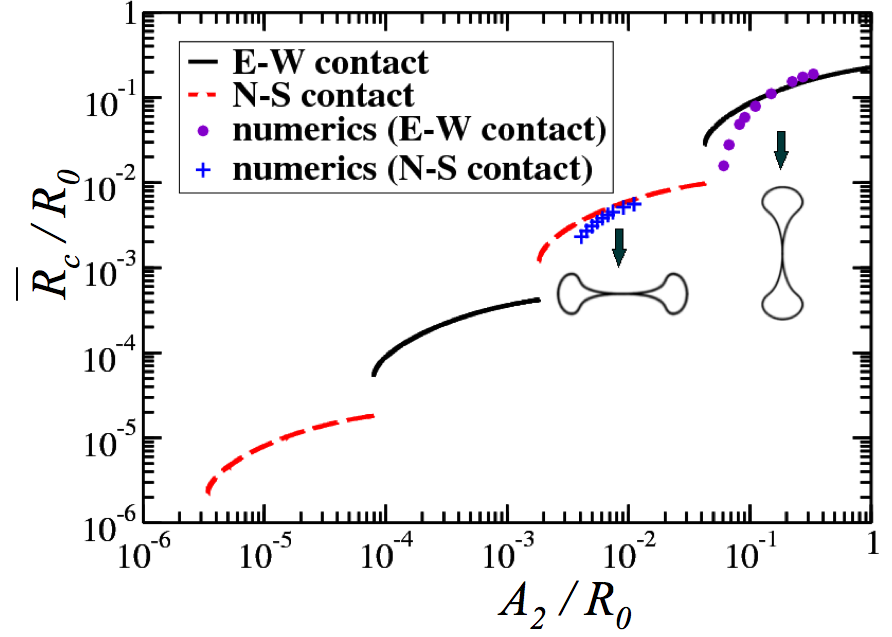}
\caption{Average radius $\bar{R}_c / R_0$ at the moment of contact as a function of the initial distortion amplitude $A_2 / R_0$. The gap in the numerical result corresponds to initial conditions that evolve into more complicated forms of self intersection.}
\end{figure}

It is worth noting that the finite-time contact is neither a physical singularity, a divergence in the velocity or the pressure, nor a mathematical singularity, corresponding to one or more singularities of the mapping function crossing the unit circle on the complex plane~\cite{shraiman84,tanveer00,lee06}.  Instead contact is a point where the mapping function becomes multivalued. 
Mathematically, the bubble break-up problem occurs within a phase space which has a natural boundary, a "wall" separating singly-connected shapes from the multiply-connected ones.  A contact corresponds to a time evolution that intersects a point on the "wall" in a finite amount of time.  
Because the time-evolution involves vibrations in the cross-section shape, both the speed and the average radius at contact has an intricate variation with respect to the initial amplitude and phase.

In conclusion,  we have conducted a numerical study of asymmetric bubble disconnection.  
The simulations show that an initial distortion excites vibrations which persist with essentially the same amplitude.  The generic outcome is a finite-time contact.  The distorted interface osculates with itself, forming a contact at a finite speed whose value depends on the initial condition. 

We thank I.~Gruzberg, N.~C.~Keim, D.~Lohse, S.~R. Nagel, L.~E.~Schmidt and P.~Wiegmann for encouragement and feedback.  This work was supported by NSF-MRSEC No.~DMR-0820054 (K.S.T.), the Keck Initiative for ultrafast imaging (University of Chicago) and NSF No.~CBET 0730629 (W.W.Z)\\
\bibliography{asymmetric4}

\begin{thebibliography}{29}
\expandafter\ifx\csname natexlab\endcsname\relax\def\natexlab#1{#1}\fi
\expandafter\ifx\csname bibnamefont\endcsname\relax
  \def\bibnamefont#1{#1}\fi
\expandafter\ifx\csname bibfnamefont\endcsname\relax
  \def\bibfnamefont#1{#1}\fi
\expandafter\ifx\csname citenamefont\endcsname\relax
  \def\citenamefont#1{#1}\fi
\expandafter\ifx\csname url\endcsname\relax
  \def\url#1{\texttt{#1}}\fi
\expandafter\ifx\csname urlprefix\endcsname\relax\def\urlprefix{URL }\fi
\providecommand{\bibinfo}[2]{#2}
\providecommand{\eprint}[2][]{\url{#2}}

\bibitem[{\citenamefont{Birkhoff and Zarantonello}(1957)}]{birkhoff57}
\bibinfo{author}{\bibfnamefont{G.}~\bibnamefont{Birkhoff}} \bibnamefont{and}
  \bibinfo{author}{\bibfnamefont{E.~H.} \bibnamefont{Zarantonello}},
  \emph{\bibinfo{title}{Jets, wakes and cavities}}
  (\bibinfo{publisher}{Academic Press}, \bibinfo{address}{New York},
  \bibinfo{year}{1957}).

\bibitem[{\citenamefont{Longuet-Higgins
  et~al.}(1991)\citenamefont{Longuet-Higgins, Kerman, and Lunde}}]{longuet91}
\bibinfo{author}{\bibfnamefont{M.~S.} \bibnamefont{Longuet-Higgins}},
  \bibinfo{author}{\bibfnamefont{B.~R.} \bibnamefont{Kerman}},
  \bibnamefont{and} \bibinfo{author}{\bibfnamefont{K.}~\bibnamefont{Lunde}},
  \bibinfo{journal}{J. Fluid Mech.} \textbf{\bibinfo{volume}{230}},
  \bibinfo{pages}{365} (\bibinfo{year}{1991}).

\bibitem[{\citenamefont{Oguz and Prosperetti}(1993)}]{oguz93}
\bibinfo{author}{\bibfnamefont{H.~N.} \bibnamefont{Oguz}} \bibnamefont{and}
  \bibinfo{author}{\bibfnamefont{A.}~\bibnamefont{Prosperetti}},
  \bibinfo{journal}{J. Fluid Mech.} \textbf{\bibinfo{volume}{257}},
  \bibinfo{pages}{111} (\bibinfo{year}{1993}).

\bibitem[{\citenamefont{Gordillo et~al.}(2005)\citenamefont{Gordillo, Sevilla,
  Rodr\'{\i}guez-Rodr\'{\i}guez, and Mart\'{\i}nez-Baz\'{a}n}}]{gordillo05}
\bibinfo{author}{\bibfnamefont{J.~M.} \bibnamefont{Gordillo}},
  \bibinfo{author}{\bibfnamefont{A.}~\bibnamefont{Sevilla}},
  \bibinfo{author}{\bibfnamefont{J.}~\bibnamefont{Rodr\'{\i}guez-Rodr\'{\i}gue%
z}}, \bibnamefont{and}
  \bibinfo{author}{\bibfnamefont{C.}~\bibnamefont{Mart\'{\i}nez-Baz\'{a}n}},
  \bibinfo{journal}{Phys. Rev. Lett.} \textbf{\bibinfo{volume}{95}},
  \bibinfo{pages}{194501} (\bibinfo{year}{2005}).

\bibitem[{\citenamefont{Burton et~al.}(2005)\citenamefont{Burton, Waldrep, and
  Taborek}}]{burton05}
\bibinfo{author}{\bibfnamefont{J.~C.} \bibnamefont{Burton}},
  \bibinfo{author}{\bibfnamefont{R.}~\bibnamefont{Waldrep}}, \bibnamefont{and}
  \bibinfo{author}{\bibfnamefont{P.}~\bibnamefont{Taborek}},
  \bibinfo{journal}{Phys. Rev. Lett.} \textbf{\bibinfo{volume}{94}},
  \bibinfo{pages}{184502} (\bibinfo{year}{2005}).

\bibitem[{\citenamefont{Keim et~al.}(2006)\citenamefont{Keim, Moller, Zhang,
  and Nagel}}]{keim06}
\bibinfo{author}{\bibfnamefont{N.~C.} \bibnamefont{Keim}},
  \bibinfo{author}{\bibfnamefont{P.}~\bibnamefont{Moller}},
  \bibinfo{author}{\bibfnamefont{W.~W.} \bibnamefont{Zhang}}, \bibnamefont{and}
  \bibinfo{author}{\bibfnamefont{S.~R.} \bibnamefont{Nagel}},
  \bibinfo{journal}{Phys. Rev. Lett.} \textbf{\bibinfo{volume}{97}},
  \bibinfo{pages}{145503} (\bibinfo{year}{2006}).

\bibitem[{\citenamefont{Bergmann et~al.}(2006)\citenamefont{Bergmann, van~der
  Meer, Stijnman, Sandtke, Prosperetti, and Lohse}}]{bergmann06}
\bibinfo{author}{\bibfnamefont{R.}~\bibnamefont{Bergmann}},
  \bibinfo{author}{\bibfnamefont{D.}~\bibnamefont{van~der Meer}},
  \bibinfo{author}{\bibfnamefont{M.}~\bibnamefont{Stijnman}},
  \bibinfo{author}{\bibfnamefont{M.}~\bibnamefont{Sandtke}},
  \bibinfo{author}{\bibfnamefont{A.}~\bibnamefont{Prosperetti}},
  \bibnamefont{and} \bibinfo{author}{\bibfnamefont{D.}~\bibnamefont{Lohse}},
  \bibinfo{journal}{Phys. Rev. Lett.} \textbf{\bibinfo{volume}{96}},
  \bibinfo{pages}{154505} (\bibinfo{year}{2006}).

\bibitem[{\citenamefont{Thoroddsen et~al.}(2007)\citenamefont{Thoroddsen, Etoh,
  and Takehara}}]{thoroddsen07}
\bibinfo{author}{\bibfnamefont{S.~T.} \bibnamefont{Thoroddsen}},
  \bibinfo{author}{\bibfnamefont{T.~G.} \bibnamefont{Etoh}}, \bibnamefont{and}
  \bibinfo{author}{\bibfnamefont{K.}~\bibnamefont{Takehara}},
  \bibinfo{journal}{Phys. Fluids} \textbf{\bibinfo{volume}{19}},
  \bibinfo{pages}{042101} (\bibinfo{year}{2007}).

\bibitem[{\citenamefont{Eggers et~al.}(2007)\citenamefont{Eggers, Fontelos,
  Leppinen, and Snoeijer}}]{egg07}
\bibinfo{author}{\bibfnamefont{J.}~\bibnamefont{Eggers}},
  \bibinfo{author}{\bibfnamefont{M.~A.} \bibnamefont{Fontelos}},
  \bibinfo{author}{\bibfnamefont{D.}~\bibnamefont{Leppinen}}, \bibnamefont{and}
  \bibinfo{author}{\bibfnamefont{J.~H.} \bibnamefont{Snoeijer}},
  \bibinfo{journal}{Phys. Rev. Lett.} \textbf{\bibinfo{volume}{98}},
  \bibinfo{pages}{094502} (\bibinfo{year}{2007}).

\bibitem[{\citenamefont{Schmidt et~al.}(2009)\citenamefont{Schmidt, Keim,
  Zhang, and Nagel}}]{schmidtpreprint}
\bibinfo{author}{\bibfnamefont{L.~E.} \bibnamefont{Schmidt}},
  \bibinfo{author}{\bibfnamefont{N.~C.} \bibnamefont{Keim}},
  \bibinfo{author}{\bibfnamefont{W.~W.} \bibnamefont{Zhang}}, \bibnamefont{and}
  \bibinfo{author}{\bibfnamefont{S.~R.} \bibnamefont{Nagel}},
  \bibinfo{journal}{Nature Phys.} \textbf{\bibinfo{volume}{5}},
  \bibinfo{pages}{343} (\bibinfo{year}{2009}).

\bibitem[{\citenamefont{Brenner et~al.}(2003)\citenamefont{Brenner,
  Hilgenfeldt, and Lohse}}]{brenner03}
\bibinfo{author}{\bibfnamefont{M.~P.} \bibnamefont{Brenner}},
  \bibinfo{author}{\bibfnamefont{S.}~\bibnamefont{Hilgenfeldt}},
  \bibnamefont{and} \bibinfo{author}{\bibfnamefont{D.}~\bibnamefont{Lohse}},
  \bibinfo{journal}{Rev. Mod. Phys.} \textbf{\bibinfo{volume}{74}},
  \bibinfo{pages}{426} (\bibinfo{year}{2003}).

\bibitem[{\citenamefont{Zeldovich and Raizer}(1966)}]{zeldovich66}
\bibinfo{author}{\bibfnamefont{Y.~B.} \bibnamefont{Zeldovich}}
  \bibnamefont{and} \bibinfo{author}{\bibfnamefont{Y.~P.}
  \bibnamefont{Raizer}}, \emph{\bibinfo{title}{Physics of Shock Waves and
  High-Temperature Hydrodynamic Phenomena}} (\bibinfo{publisher}{Dover},
  \bibinfo{address}{Mineola New York}, \bibinfo{year}{1966}).

\bibitem[{\citenamefont{Shi et~al.}(1994)\citenamefont{Shi, Brenner, and
  Nagel}}]{shi94}
\bibinfo{author}{\bibfnamefont{X.~D.} \bibnamefont{Shi}},
  \bibinfo{author}{\bibfnamefont{M.~P.} \bibnamefont{Brenner}},
  \bibnamefont{and} \bibinfo{author}{\bibfnamefont{S.~R.} \bibnamefont{Nagel}},
  \bibinfo{journal}{Science} \textbf{\bibinfo{volume}{265}},
  \bibinfo{pages}{219} (\bibinfo{year}{1994}).

\bibitem[{\citenamefont{Eggers}(1997)}]{egg97}
\bibinfo{author}{\bibfnamefont{J.}~\bibnamefont{Eggers}},
  \bibinfo{journal}{Rev. Mod. Phys.} \textbf{\bibinfo{volume}{69}},
  \bibinfo{pages}{865} (\bibinfo{year}{1997}).

\bibitem[{\citenamefont{Doshi et~al.}(2003)\citenamefont{Doshi, Cohen, Zhang,
  Siegel, Howell, Basaran, and Nagel}}]{doshi03}
\bibinfo{author}{\bibfnamefont{P.}~\bibnamefont{Doshi}},
  \bibinfo{author}{\bibfnamefont{I.}~\bibnamefont{Cohen}},
  \bibinfo{author}{\bibfnamefont{W.~W.} \bibnamefont{Zhang}},
  \bibinfo{author}{\bibfnamefont{M.}~\bibnamefont{Siegel}},
  \bibinfo{author}{\bibfnamefont{P.}~\bibnamefont{Howell}},
  \bibinfo{author}{\bibfnamefont{O.~A.} \bibnamefont{Basaran}},
  \bibnamefont{and} \bibinfo{author}{\bibfnamefont{S.~R.} \bibnamefont{Nagel}},
  \bibinfo{journal}{Science} \textbf{\bibinfo{volume}{302}},
  \bibinfo{pages}{1185} (\bibinfo{year}{2003}).

\bibitem[{\citenamefont{Whitham}(1957)}]{whitham57}
\bibinfo{author}{\bibfnamefont{G.~B.} \bibnamefont{Whitham}},
  \bibinfo{journal}{J. Fluid Mech.} \textbf{\bibinfo{volume}{2}},
  \bibinfo{pages}{145} (\bibinfo{year}{1957}).

\bibitem[{\citenamefont{Plesset and Prosperetti}(1977)}]{plesset77}
\bibinfo{author}{\bibfnamefont{M.~S.} \bibnamefont{Plesset}} \bibnamefont{and}
  \bibinfo{author}{\bibfnamefont{A.}~\bibnamefont{Prosperetti}},
  \bibinfo{journal}{Ann. Rev. Fluid Mech.} \textbf{\bibinfo{volume}{9}},
  \bibinfo{pages}{145} (\bibinfo{year}{1977}).

\bibitem[{\citenamefont{Evans}(1996)}]{evans96}
\bibinfo{author}{\bibfnamefont{A.~K.} \bibnamefont{Evans}},
  \bibinfo{journal}{Phys. Rev. E} \textbf{\bibinfo{volume}{54}},
  \bibinfo{pages}{5004} (\bibinfo{year}{1996}).

\bibitem[{\citenamefont{Plewa et~al.}(2004)\citenamefont{Plewa, Calder, and
  Lamb}}]{plewa04}
\bibinfo{author}{\bibfnamefont{T.}~\bibnamefont{Plewa}},
  \bibinfo{author}{\bibfnamefont{A.~C.} \bibnamefont{Calder}},
  \bibnamefont{and} \bibinfo{author}{\bibfnamefont{D.~G.} \bibnamefont{Lamb}},
  \bibinfo{journal}{Astrophys. J.} \textbf{\bibinfo{volume}{612}},
  \bibinfo{pages}{L37} (\bibinfo{year}{2004}).

\bibitem[{\citenamefont{Maeda et~al.}(2008)\citenamefont{Maeda, Kawabata,
  Mazzali, Tanaka, Valenti, Nomoto, Hattori, Deng, Pian, Taubenberger
  et~al.}}]{maeda08}
\bibinfo{author}{\bibfnamefont{K.}~\bibnamefont{Maeda}},
  \bibinfo{author}{\bibfnamefont{K.}~\bibnamefont{Kawabata}},
  \bibinfo{author}{\bibfnamefont{P.~A.} \bibnamefont{Mazzali}},
  \bibinfo{author}{\bibfnamefont{M.}~\bibnamefont{Tanaka}},
  \bibinfo{author}{\bibfnamefont{S.}~\bibnamefont{Valenti}},
  \bibinfo{author}{\bibfnamefont{K.}~\bibnamefont{Nomoto}},
  \bibinfo{author}{\bibfnamefont{T.}~\bibnamefont{Hattori}},
  \bibinfo{author}{\bibfnamefont{J.}~\bibnamefont{Deng}},
  \bibinfo{author}{\bibfnamefont{E.}~\bibnamefont{Pian}},
  \bibinfo{author}{\bibfnamefont{S.}~\bibnamefont{Taubenberger}},
  \bibnamefont{et~al.}, \bibinfo{journal}{Science}
  \textbf{\bibinfo{volume}{319}}, \bibinfo{pages}{1220} (\bibinfo{year}{2008}).

\bibitem[{\citenamefont{Zwaan et~al.}(2007)\citenamefont{Zwaan, Gac, Tsuji, and
  Ohl}}]{zwaan07}
\bibinfo{author}{\bibfnamefont{E.}~\bibnamefont{Zwaan}},
  \bibinfo{author}{\bibfnamefont{S.~L.} \bibnamefont{Gac}},
  \bibinfo{author}{\bibfnamefont{K.}~\bibnamefont{Tsuji}}, \bibnamefont{and}
  \bibinfo{author}{\bibfnamefont{C.-D.} \bibnamefont{Ohl}},
  \bibinfo{journal}{Phys. Rev. Lett.} \textbf{\bibinfo{volume}{98}},
  \bibinfo{eid}{254501} (\bibinfo{year}{2007}).

\bibitem[{\citenamefont{Schmidt}(2008)}]{schmidtThesis}
\bibinfo{author}{\bibfnamefont{L.~E.} \bibnamefont{Schmidt}}, Ph.D. thesis,
  \bibinfo{school}{University of Chicago} (\bibinfo{year}{2008}).

\bibitem[{\citenamefont{Landau and Lifshitz}(1987)}]{landaubook}
\bibinfo{author}{\bibfnamefont{L.~D.} \bibnamefont{Landau}} \bibnamefont{and}
  \bibinfo{author}{\bibfnamefont{E.~M.} \bibnamefont{Lifshitz}},
  \emph{\bibinfo{title}{Fluid Mechanics}}
  (\bibinfo{publisher}{Butterworth-Heinemann}, \bibinfo{address}{Oxford},
  \bibinfo{year}{1987}).

\bibitem[{\citenamefont{Dyachenko et~al.}(1996)\citenamefont{Dyachenko,
  Kuznetsov, Spector, and Zakharov}}]{dyachenko96}
\bibinfo{author}{\bibfnamefont{A.~I.} \bibnamefont{Dyachenko}},
  \bibinfo{author}{\bibfnamefont{E.~A.} \bibnamefont{Kuznetsov}},
  \bibinfo{author}{\bibfnamefont{M.~D.} \bibnamefont{Spector}},
  \bibnamefont{and} \bibinfo{author}{\bibfnamefont{V.~E.}
  \bibnamefont{Zakharov}}, \bibinfo{journal}{Phys. Lett. A}
  \textbf{\bibinfo{volume}{221}}, \bibinfo{pages}{73} (\bibinfo{year}{1996}).

\bibitem[{\citenamefont{Zakharov et~al.}(2002)\citenamefont{Zakharov,
  Dyachenko, and Vasilyev}}]{zakharov02}
\bibinfo{author}{\bibfnamefont{V.~E.} \bibnamefont{Zakharov}},
  \bibinfo{author}{\bibfnamefont{A.~I.} \bibnamefont{Dyachenko}},
  \bibnamefont{and} \bibinfo{author}{\bibfnamefont{O.~A.}
  \bibnamefont{Vasilyev}}, \bibinfo{journal}{Europ. J. Mech. B}
  \textbf{\bibinfo{volume}{21}}, \bibinfo{pages}{283} (\bibinfo{year}{2002}).

\bibitem[{\citenamefont{Shraiman and Bensimon}(1984)}]{shraiman84}
\bibinfo{author}{\bibfnamefont{B.}~\bibnamefont{Shraiman}} \bibnamefont{and}
  \bibinfo{author}{\bibfnamefont{D.}~\bibnamefont{Bensimon}},
  \bibinfo{journal}{Phys. Rev. A} \textbf{\bibinfo{volume}{30}},
  \bibinfo{pages}{2840} (\bibinfo{year}{1984}).

\bibitem[{\citenamefont{Carrier et~al.}(1966)\citenamefont{Carrier, Krook, and
  Pearson}}]{carrier}
\bibinfo{author}{\bibfnamefont{G.~F.} \bibnamefont{Carrier}},
  \bibinfo{author}{\bibfnamefont{M.}~\bibnamefont{Krook}}, \bibnamefont{and}
  \bibinfo{author}{\bibfnamefont{C.~E.} \bibnamefont{Pearson}},
  \emph{\bibinfo{title}{Functions of a complex variable}}
  (\bibinfo{publisher}{McGraw-Hill}, \bibinfo{address}{New York},
  \bibinfo{year}{1966}).

\bibitem[{\citenamefont{Tanveer}(2000)}]{tanveer00}
\bibinfo{author}{\bibfnamefont{S.}~\bibnamefont{Tanveer}}, \bibinfo{journal}{J.
  Fluid Mech.} \textbf{\bibinfo{volume}{409}}, \bibinfo{pages}{273}
  (\bibinfo{year}{2000}).

\bibitem[{\citenamefont{Lee et~al.}(2006)\citenamefont{Lee, Bettelheim, and
  Wiegmann}}]{lee06}
\bibinfo{author}{\bibfnamefont{S.~Y.} \bibnamefont{Lee}},
  \bibinfo{author}{\bibfnamefont{E.}~\bibnamefont{Bettelheim}},
  \bibnamefont{and} \bibinfo{author}{\bibfnamefont{P.}~\bibnamefont{Wiegmann}},
  \bibinfo{journal}{Physica D} \textbf{\bibinfo{volume}{219}},
  \bibinfo{pages}{22} (\bibinfo{year}{2006}).

\end{thebibliography}
\end{document}